\documentclass[a4paper,11pt]{article}
\usepackage{jinstpub} % for details on the use of the package, please see the JINST-author-manual
\usepackage{lineno}
\usepackage{subcaption}
\graphicspath{{Figures/}}
\usepackage{xfrac}

\title{Integration Concept of the CBM Micro Vertex Detector}

\author[a,1]{Franz Matejcek\note{Corresponding author.}}
\author[c]{, Ali-Murteza Altingun}
\author[a]{, Julio Andary}
\author[a]{, Benedict Arnoldi-Meadows}
\author[c]{, Jerome Baudot}
\author[c]{, Gregory Bertolone}
\author[c]{, Auguste Besson}
\author[a]{, Norbert Bialas}
\author[a]{, Christopher Braun}
\author[c]{, Roma Bugiel}
\author[c]{, Gilles Claus}
\author[c]{, Claude Colledani}
\author[a,c]{, Hasan Darwish}
\author[a,b]{, Michael Deveaux}
\author[c]{, Andrei Dorokhov}
\author[c]{, Guy Dozi\`ere}
\author[c]{, Ziad El Bitar}
\author[a,b]{, Ingo Fröhlich}
\author[c]{, Mathieu Goffe}
\author[a]{, Benedikt Gutsche}
\author[c]{, Abdelkader Himmi}
\author[c]{, Christine Hu-Guo}
\author[c]{, Kimmo Jaaskelainen}
\author[f]{, Oliver Keller}
\author[a]{, Michal Koziel}
\author[a]{, Jan Michel}
\author[c]{, Frederic Morel}
\author[a]{, Christian M\"untz}
\author[c]{, Hung Pham}
\author[b]{, Christian Joachim Schmidt}
\author[a]{, Stefan Schreiber}
\author[c]{, Matthieu Specht}
\author[a,b,d]{, Joachim Stroth}
\author[a]{, Eva-dhidho Taka}
\author[c]{, Isabelle Valin}
\author[a]{, Roland Weirich}
\author[c]{, Y\"ue Zhao}
\author[e]{and Marc Winter}
\affiliation[a]{Institut für Kernphysik, Goethe-Universität Frankfurt, Max-von-Laue-Straße 1, 60438 Frankfurt (Germany)}
\affiliation[b]{GSI Helmholtzzentrum für Schwerionenforschung GmbH, Planckstraße 1, 64291 Darmstadt (Germany)}
\affiliation[c]{Universit\'e de Strasbourg, CNRS, IPHC UMR 7178, 67037 Strasbourg (France)}

\affiliation[d]{Helmholtz Forschungsakademie Hessen für FAIR, Max-von-Laue-Straße 12, 60438 Frankfurt (Germany)}
\affiliation[e]{Universit\'e Paris-Saclay, CNRS/IN2P3, IJCLab, 91405 Orsay (France)}
\affiliation[f]{Facility for Antiproton and Ion Research GmbH, Planckstraße 1, 64291 Darmstadt (Germany)}

\emailAdd{matejcek@physik.uni-frankfurt.de}

\abstract{The Micro Vertex Detector (MVD) is the most upstream detector of the fixed-target Compressed Baryonic Matter Experiment (CBM) at the future Facility for Antiproton and Ion Research (FAIR). It enables high-precision low-momentum tracking in direct proximity of the target. Reaching the stringent requirements for the MVD, a material budget of~$0.3\,-\,0.5\%\,X_0$ per layer, operating the dedicated CMOS MAPS~(`MIMOSIS') in the target vacuum, the strong magnetic dipole field, and a harsh radiation environment~(5\,Mrad, $7\times10^{13}\,n_{\text{eq}}/\text{cm}^2$ per CBM year), poses an unprecedented integration challenge.\par
In this paper, the integration concept of the detector is be outlined, elaborating on the selection and preparation of materials, assembly procedures, and quality assessment steps in the ongoing preparation of pre-series production and detector commissioning in 2028.
}


\begin{document}
\maketitle
\flushbottom
\section{Introduction}
\paragraph{CBM} The Compressed Baryonic Matter Experiment (CBM)~\cite{CBM:2016kpk} is currently being set up at the future Facility for Antiproton and Ion Research (FAIR) to study strongly interacting matter at neutron star core densities in fixed-target Au-Au collisions ($\sqrt{s_\text{NN}}=2.9\,-\,4.9$\,GeV; beam-target interaction rate up to $10^7$\,MHz; polar angular acceptance of 2.5$^{\circ}\le\theta\le25^{\circ}$ over the full azimuth).\par
\paragraph{MVD} The CMOS-MAPS-based Micro Vertex Detector (MVD)~\cite{MVD:TDR} is placed in closest proximity to the fixed target and, together with the micro-strip-based Silicon Tracking System (STS), allows to reconstruct tracks down to a few 100\,MeV/c in a high track-density environment. The MVD
\begin{itemize}
\item operates in the moderate target vacuum and in a strong magnetic dipole field $\left(\int B\cdot dz=1\,\text{T}\text{m}\right)$ while featuring a material budget between $0.3\,-\,0.5\%\,X_0$ per layer.
\vspace{-8pt}
\item comprises four planar tracking layers, between 8\,cm and 20\,cm downstream (5.4\,mm from the beam axis), and is equipped with 288 CMOS MAPS~(`MIMOSIS', designed for the MVD) that feature a power density of up to 75\,mW/cm$^2$ and cover $\approx$\,0.15\,m$^2$ with $\approx$\,150\,M pixels.
\vspace{-8pt}
\item features a single-point spatial measuring precision of 5\,\textmu m in a free-streaming readout ($t_{\text{Frame}}=5$\,\textmu s), resulting in a secondary vertexing precision of $\sigma_z\approx70$\,\textmu m.
\vspace{-8pt}
\item employs a hit rate capability of 20~(80)\,MHz/cm$^2$~mean\,(peak) to cope with beam-target interaction rates up 0.1\,MHz Au+Au, 10\,MHz p+Au, such that the most exposed sensors accumulate the end-of-lifetime (EOL) dose (5\,Mrad ionizing and $7\times10^{13}\,n_{\text{eq}}/\text{cm}^2$ non-ionizing) already after one~CBM year, \textit{i.\,e.}~two~months of beamtime.
\end{itemize}

\paragraph{Conceptual Design} The design of the MVD is driven by optimized material budget and hence, minimized multiple scattering of low-momentum particles in the CBM energy regime. The interplay of active area, mechanical stability and thermal performance with material budget constraints, while ensuring vacuum compatibility, results in large, highly integrated modules with 8 to 28~sensors per module. A module represents a quadrant of a station and two modules form a Half-Station.\par
\begin{figure}[ht]
    \centering
    \begin{minipage}{.46\linewidth}
        \begin{subfigure}[t]{\linewidth}
        \hspace{-12pt}
            \includegraphics[angle=270, width=1.1\textwidth]{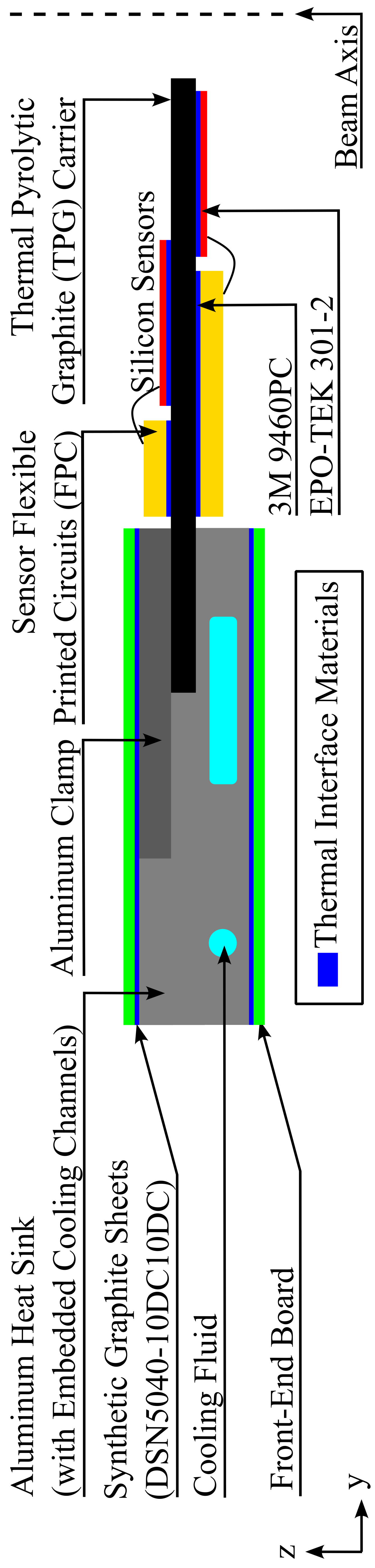}
            \caption{}
            \label{Fig:Module_CrossSection}
        \end{subfigure} \\
        \begin{subfigure}[b]{\linewidth}
        \centering
            \includegraphics[height=4.8cm]{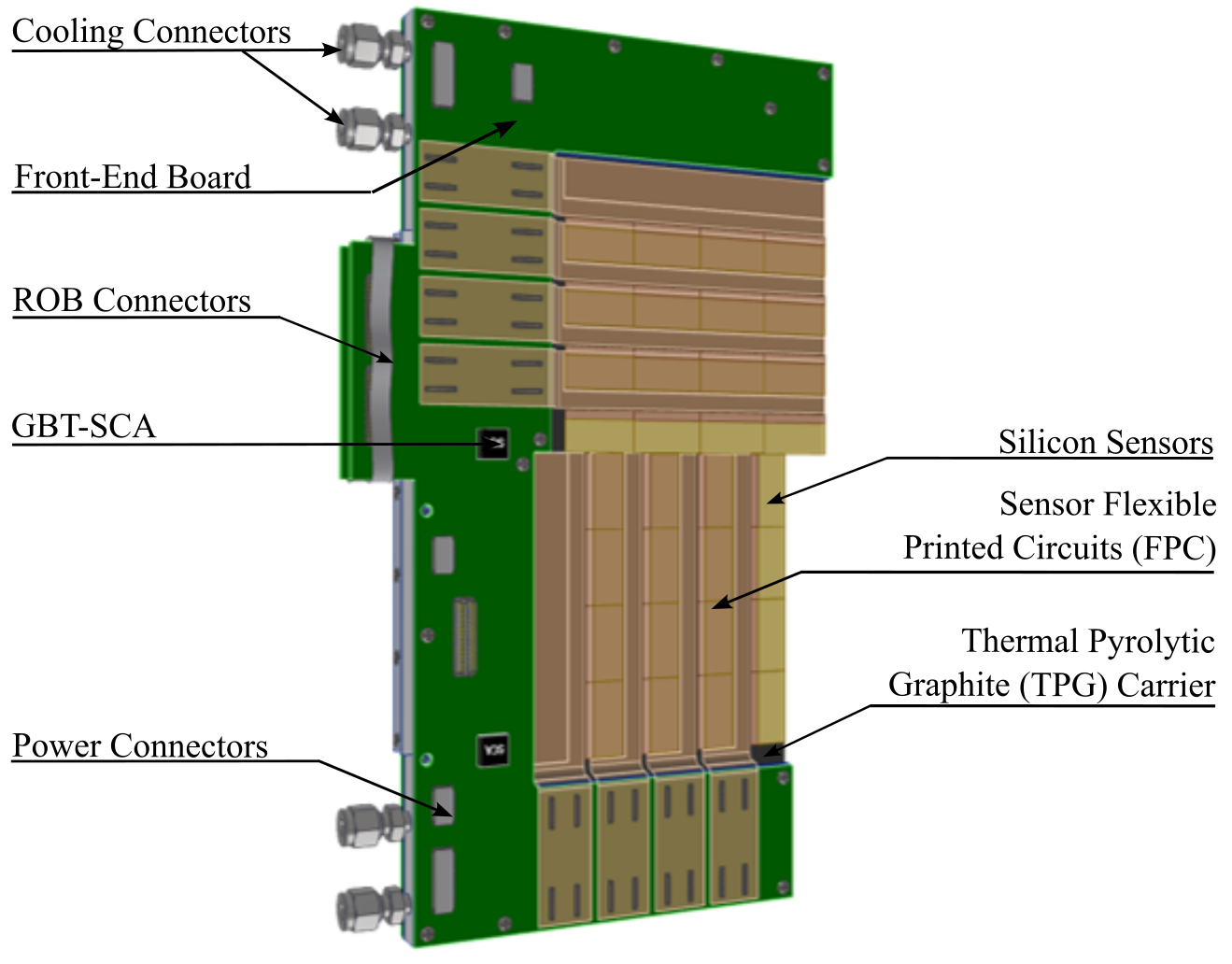}
            \caption{}
            \label{Fig:CAD_HalfStation}
        \end{subfigure} 
    \end{minipage}
    \hspace{0.5cm}
    \begin{minipage}{.42\linewidth}
            \begin{subfigure}[t]{\linewidth}
                \includegraphics[width=\textwidth]{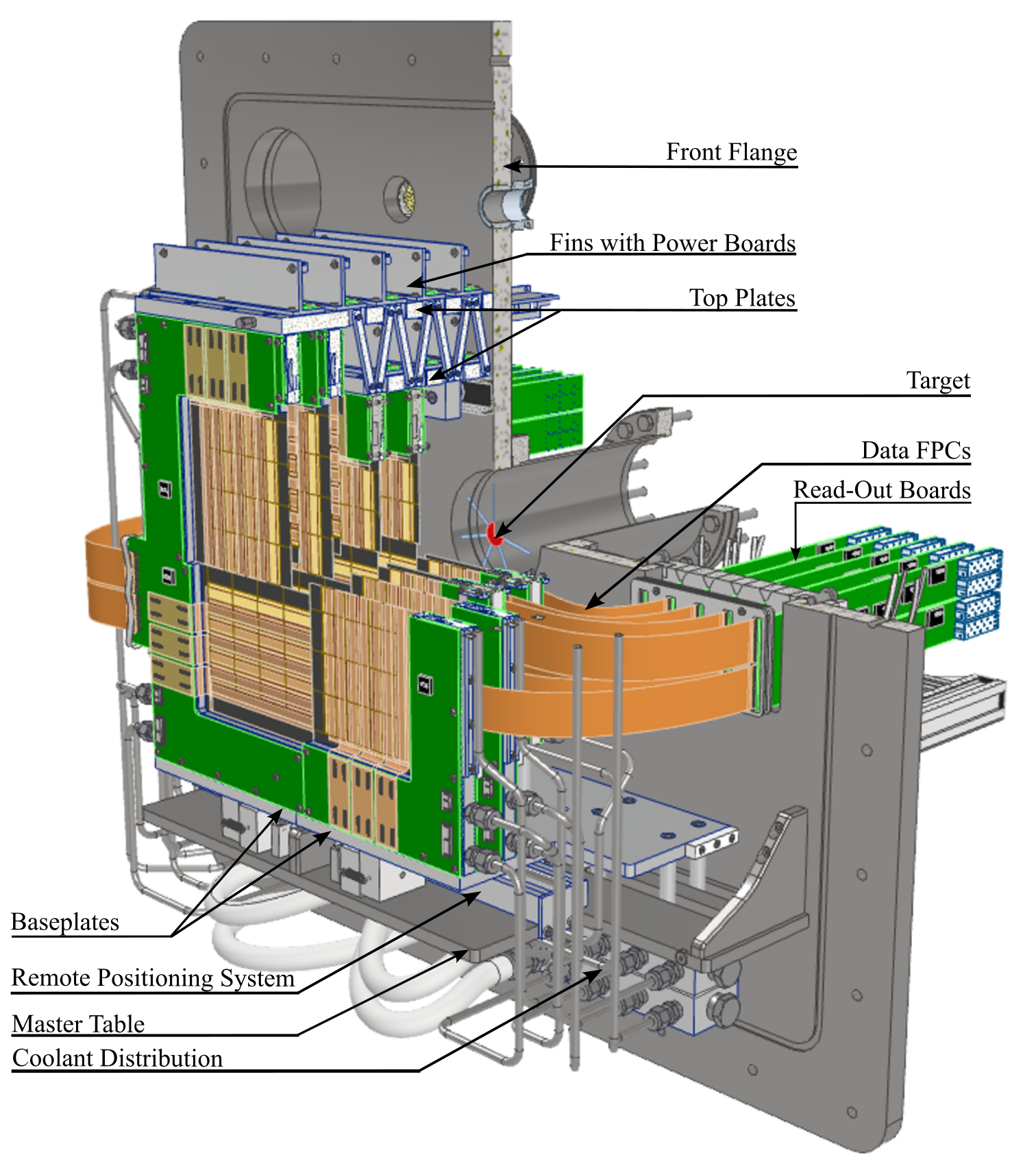}
                \caption{}
                \label{Fig:CAD_MVD}
            \end{subfigure}
        \end{minipage}
    \caption{Illustration showing the cross section of a module (not
to scale)~(\subref{Fig:Module_CrossSection}). CAD rendering of a Half-Station~(\subref{Fig:CAD_HalfStation}). CAD rendering of a \sfrac{3}{4}-section of the MVD~(\subref{Fig:CAD_MVD}).}
\end{figure}

The power dissipated by the sensors ($\approx$70\,W, in vacuum) is conducted through 380\,\textmu m thin Thermal Pyrolytic Graphite (TPG) sheets in the acceptance and neutralized in actively cooled Heat Sinks in the periphery. This minimizes material in the acceptance while keeping a low temperature gradient $<$\,10\,K over the TPG. Sensors are placed on both sides of the carrier to guarantee a 100\% fill factor, while accounting for a passive, digital part occupying $\approx$20\,\% of the sensor area. They are powered, steered, and read out over thin Flexible Printed Circuits (FPCs) that connect the Front-End Electronics mounted on the Heat Sinks. Figure~\ref{Fig:Module_CrossSection} shows the cross-section of MVD modules, figure~\ref{Fig:CAD_HalfStation} shows a corresponding CAD rendering of a Half-Station. The MVD comprises two types of Half-Stations, \textit{i.\,e.} ones with 16 and 54 sensors for the first two and last two stations, respectively.\par
\vspace{-7pt}
Modules are assembled to Half-Stations and then to two Half-Detectors which can be moved between measuring position and a safe position 5\,cm lateral from the beam axis by a Remote Positioning System to prevent direct beam impact \textit{e.\,g.} during beam tuning. All parts are mounted onto a stainless steel Front Flange which is mounted onto the target box. Low voltage is provided by FEASTMP~DC/DC-converter modules inside the vacuum. The CBM-compatible readout is based on GBTx and VTRx/VTTx chips placed on dedicated Read-Out Boards (ROBs) outside the vacuum with detector controls enabled by GBT-SCA chips placed on the Front-End Boards (FEBs)~\cite{CBM:DAQ}. Figure~\ref{Fig:CAD_MVD} shows a CAD rendering of the MVD.\par
\vspace{-5pt} 
\section{Electrical Integration}
\begin{figure}[ht]
    \centering
    \begin{minipage}{.59\linewidth}
        \begin{subfigure}[t]{\linewidth}
            \centering
        \includegraphics[height=4cm]{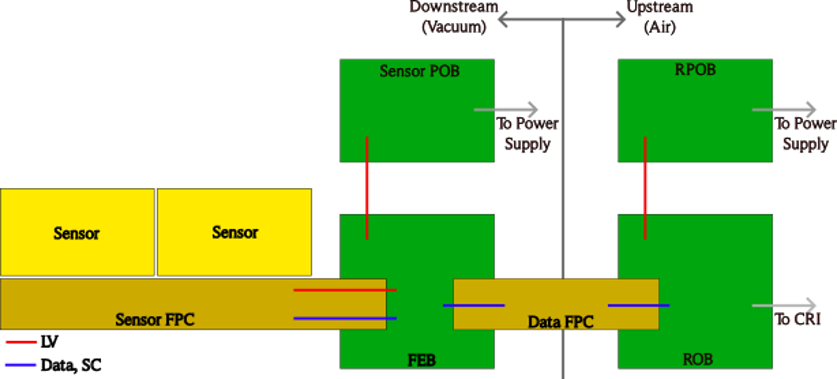}
            \caption{}
            \label{Fig:Electrical_Scheme}
        \end{subfigure}
    \end{minipage}
    \hfill
    \begin{minipage}{.39\linewidth}
        \begin{subfigure}[t]{\linewidth}
            \centering
            \includegraphics[trim = 700pt 250pt 100pt 400pt, clip,height=4cm]{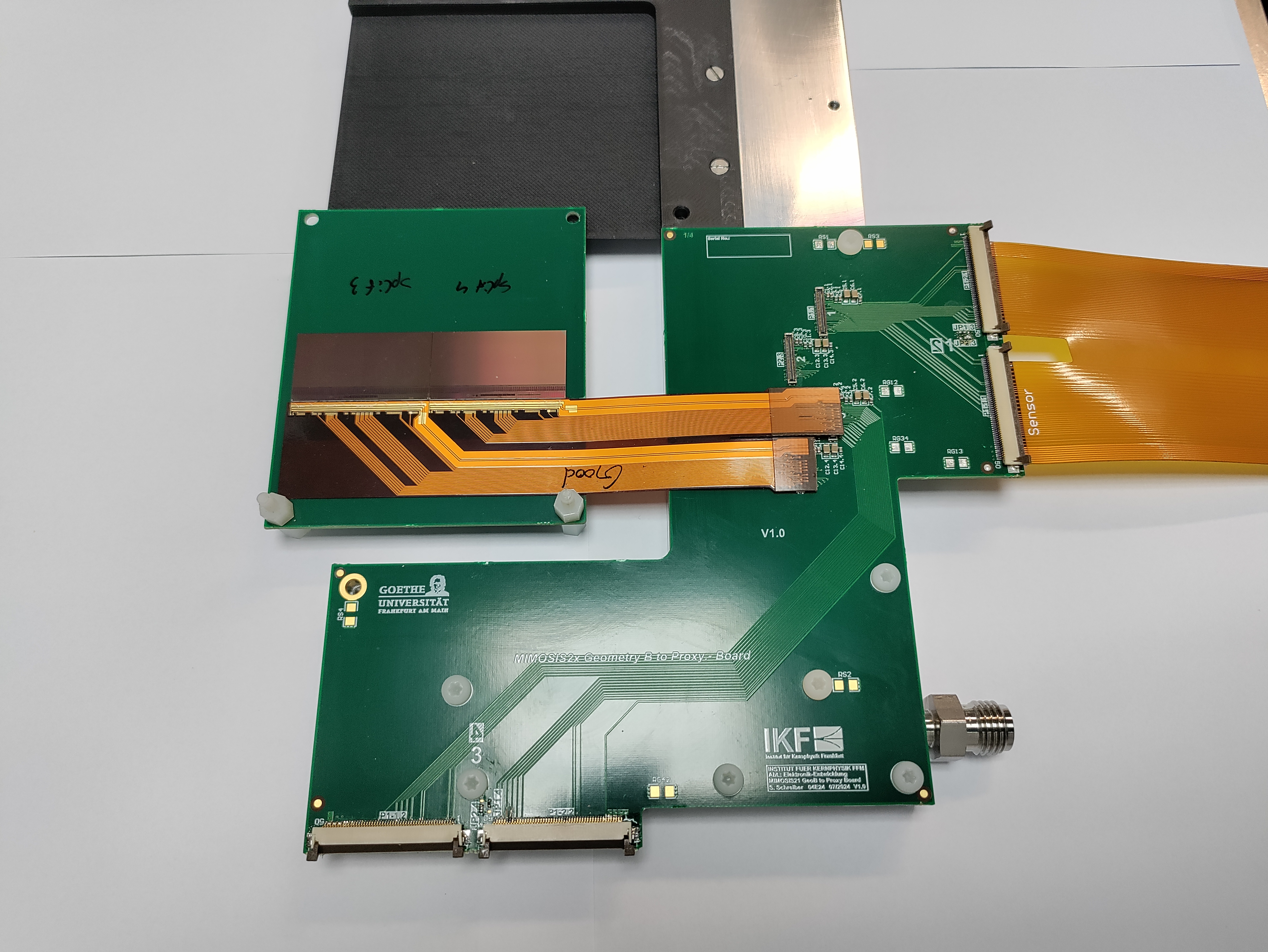}
            \caption{}
            \label{Fig:MVD_FPC}
        \end{subfigure}
        \end{minipage}
    \caption{Illustration showing the powering, slow control and readout scheme (not to scale)~(\subref{Fig:Electrical_Scheme}). The first MVD Sensor FPC prototype with Front-End-Board mounted on a Heat Sink~(\subref{Fig:MVD_FPC}).}
\end{figure}

Figure~\ref{Fig:Electrical_Scheme} shows a schematic overview of the powering and readout system of the MVD with components mounted in- and outside the vacuum box.
\paragraph{Sensor FPC} 144 single-layer, ultra-thin FPCs based on 12\,\textmu m Cu-traces ($\approx0.05\%\,X_0$\footnote{R\&D on aluminum FPCs reducing the material budget further is ongoing in parallel together with LTU~Ltd.~Kharkiv.}) connect two sensors in parallel with wire bonds to the low-voltage (LV) and biasing voltage supply, the slow control, clock, and the readout. First prototypes are currently being tested, see figure~\ref{Fig:MVD_FPC}.
\paragraph{Front-End Board} 32 PCBs mounted on both sides of the Heat Sinks interface the electrical services and additionally features passive filtering elements for the LV and biasing lines, and a GBT-SCA providing I$^2$C communication with the sensors and ADC channels for detector controls.\par
\vspace{-3pt}
\paragraph{Data FPC} 40 three-layer, impedance-controlled, shielded FPCs connect the FEBs with the ROBs, routing the global 40\,MHz CBM~clock and the data links.
\vspace{-3pt}
\paragraph{Read-Out Board} 40 ROBs, comprising three GBTx, one GBT-SCA, one VTRx and VTTx, placed outside the vacuum receive the data of 844 320\,Mbit/s data links and convert it from electrical to optical signals. The data is sent to the FPGA-based CBM Common-Readout-Interface (CRI)~\cite{CBM:DAQ}. 
\vspace{-3pt}
\paragraph{Power Board} 26 PCBs convert the 12\,V input voltage of the detector, providing 1.8\,V for the sensors (digital and analog), 1.5\,V for the GBT-SCA and GBTx, and 2.5\,V for the VTRx and VTTx.

\vspace{-11pt}

\section{Mechanical Integration}
\paragraph{Sensor}
The sensing element of the MVD is a Monolithic Active Pixel Sensor (MAPS) called MIMOSIS which has been derived from ALPIDE~\cite{ALPIDE}. It is developed by IPHC Strasbourg and fabricated in the TowerJazz~180\,nm CMOS Imaging Process. It features a fully depleted sensing node with in-pixel amplification, shaping and discrimination, on-matrix clustering, and a priority-encoder circuit with a re-designed digital front- and back-end based on ALPIDE with higher bandwidth and an elastic buffer. The MIMOSIS form factor is 31.15\,$\times$\,17.25\,mm$^2$, comprising a matrix of 1024\,$\times$\,504 pixels (26.88\,$\times$\,30.24\,\textmu m$^2$). The next-to-final generation sensor, MIMOSIS-2.1, features a standard (ALPIDE) and a p-stop modified epitaxial layer, with either 50\,\textmu m or 25\,\textmu m thickness, and DC- and AC-coupled pixels. Sensors are thinned to 70\,(50)\,\textmu m for 50\,(25)\,\textmu m epitaxial layer. The validation of the pixel options is ongoing, finalizing the pixel choice and preparing for the final sensor submission in 2025 with the current status summarized in~\cite{MVD:MIMOSIS_Performance}.\par
\begin{figure}[ht]
    \centering
    \begin{minipage}{.49\linewidth}
        \begin{subfigure}[t]{\linewidth}
            \centering
            \includegraphics[height=3.7cm]{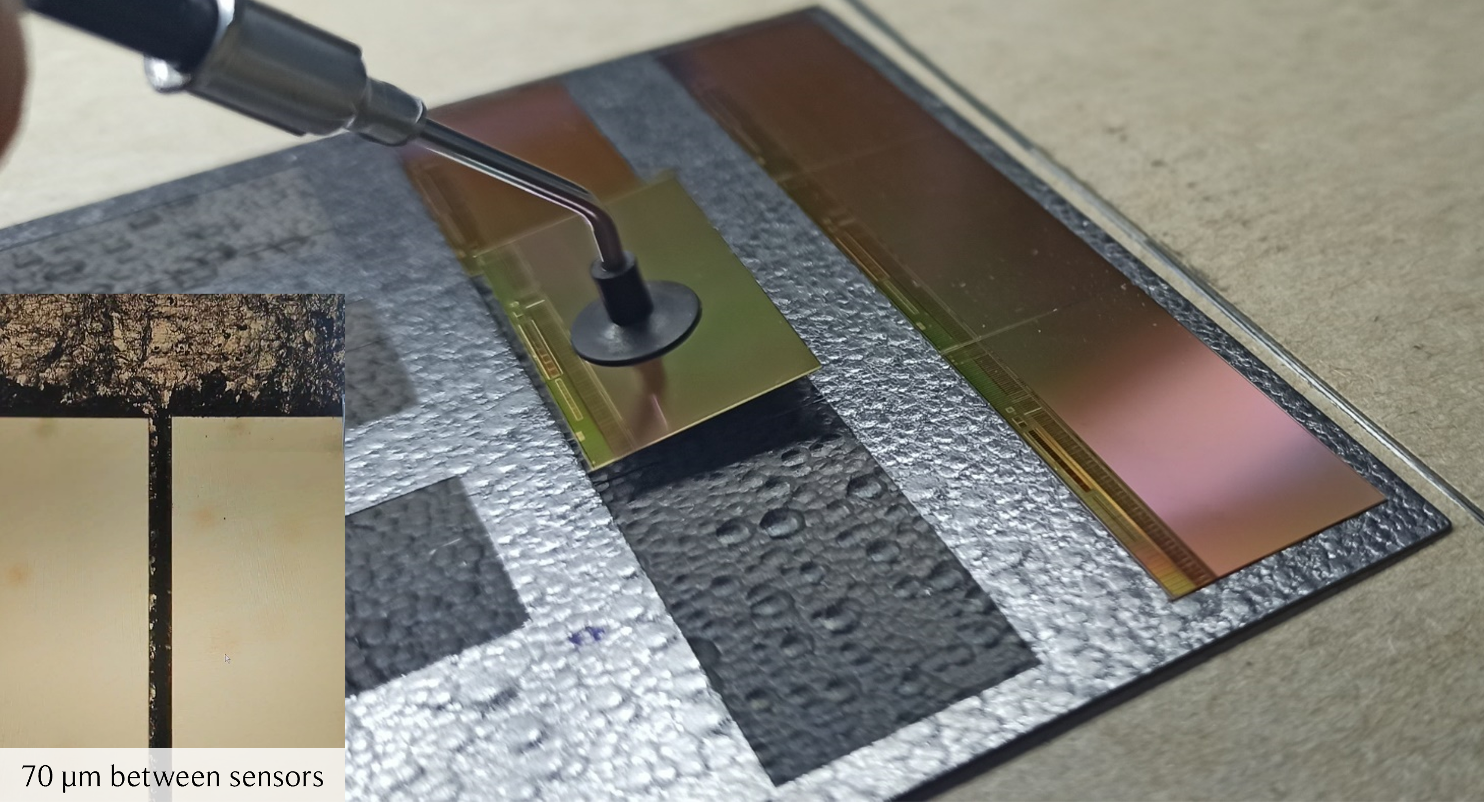}
            \caption{}
            \label{Fig:TPG_Hatches}
        \end{subfigure}
    \end{minipage}
    \hfill
    \begin{minipage}{.49\linewidth}
        \begin{subfigure}[t]{\linewidth}
            \centering
            \includegraphics[height=3.7cm]{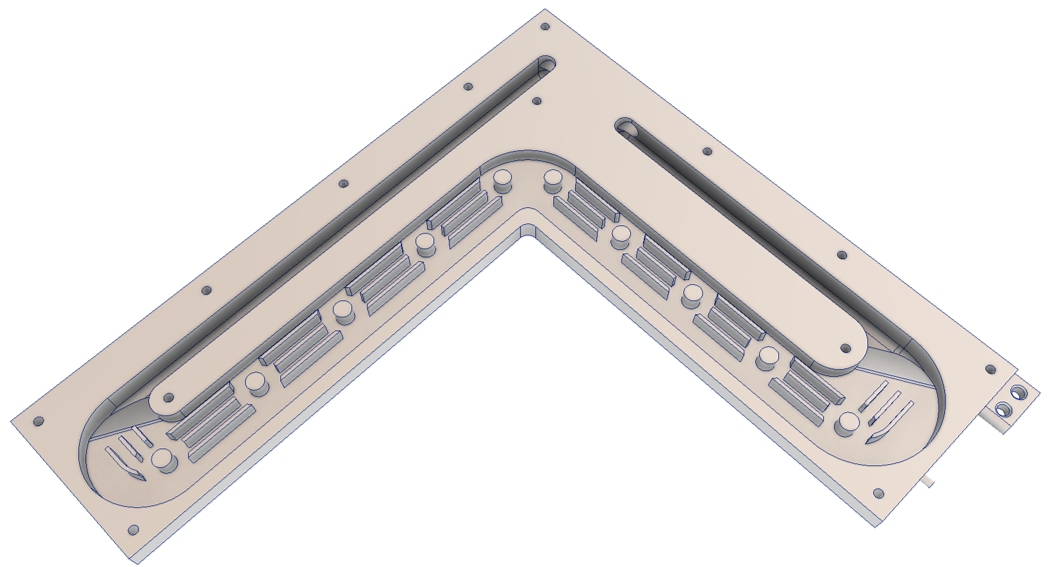}
            \caption{}
            \label{Fig:HeatSink}
        \end{subfigure}
    \end{minipage}
    \caption{Thinned MIMOSIS sensors placed in hatches on a TPG carrier~(\subref{Fig:TPG_Hatches}). CAD view of the Heat Sink's cooling channel~(\subref{Fig:HeatSink})~\cite{MVD:Cooling}.}
\end{figure}

Quality is assessed by single-die probe testing which is currently optimized with MIMOSIS-2.1. The automated test procedure includes monitoring of the analog, digital and biasing currents, sweeping of the back bias, register addressing, reading and writing, tests of Chip IDs, functionality tests of the eight high-speed data outputs, scans of the steering DACs, tests of the self-consistency of the data words, a threshold scan by S-curves, and a scan of noisy and dead pixels.
\paragraph{Carrier}\label{Ch:Carrier} 380\,\textmu m thin Thermal Pyrolytic Graphite (TPG) sheets\footnote{Integration on Carriers of 250\,\textmu m thin TPG sheets reducing the material budget further is exercised in parallel.} ($\lambda_{\text{in plane}}\geq\,$1500\,W/m/K) serve as mechanical support and cooling interface. The TPG is prepared for integration first by polishing it, reducing the surface roughness to few 10\,\textmu m. Afterwards, the sheets are cut to a rectangular shape. Deepenings ($\approx50$\,\textmu m deep hatches) to place the sensors in are cut into the surface with a UV-laser ablation device, see figure~\ref{Fig:TPG_Hatches}. Together with fiducial marks on the TPG, these hatches allow for precise ($\approx10$\,\textmu m) alignment and placement of sensors on both sides of the Carrier and an integration without jigs. Next, the TPG is coated with 5\,\textmu m of parylene-C to eliminate the surface's abrasiveness and electrical conductivity. Lastly, the surface is plasma activated to allow for gluing sensors with Epo-Tek\,301-2 (curing 48\,h at 23\,$^{\circ}$C) and Sensor FPCs with 3M\textsuperscript{TM} VHB\textsuperscript{TM} tape~9460.\par
\vspace{-3pt}
\paragraph{Heat Sink} Aluminum Heat Sinks represent the mechanical detector frame outside the acceptance and neutralize the dissipated power in a CNC-milled cooling channel optimized for high heat transfer with low pressure drops with mono-phase liquid 3M\textsuperscript{TM} Novec\textsuperscript{TM}~649 at $-20\,^{\circ}$C\footnote{Increased operation temperatures and water-based alternatives to cope with the EU-wide PFAS use restrictions are under investigation.}, see figure~\ref{Fig:HeatSink}. The cooling system is designed for a wide temperature range above $+30\,^{\circ}$C and various coolants including water and water-based mediums. Details on the cooling performance can be found in~\cite{MVD:Cooling}.\par
\vspace{-11pt}

\section{Assembly Techniques}
\begin{figure}[ht]
    \centering
    \begin{minipage}{.52\linewidth}
        \begin{subfigure}[t]{\linewidth}
            \centering
            \includegraphics[trim=60pt 150pt 60pt 150pt, clip, height=5cm]{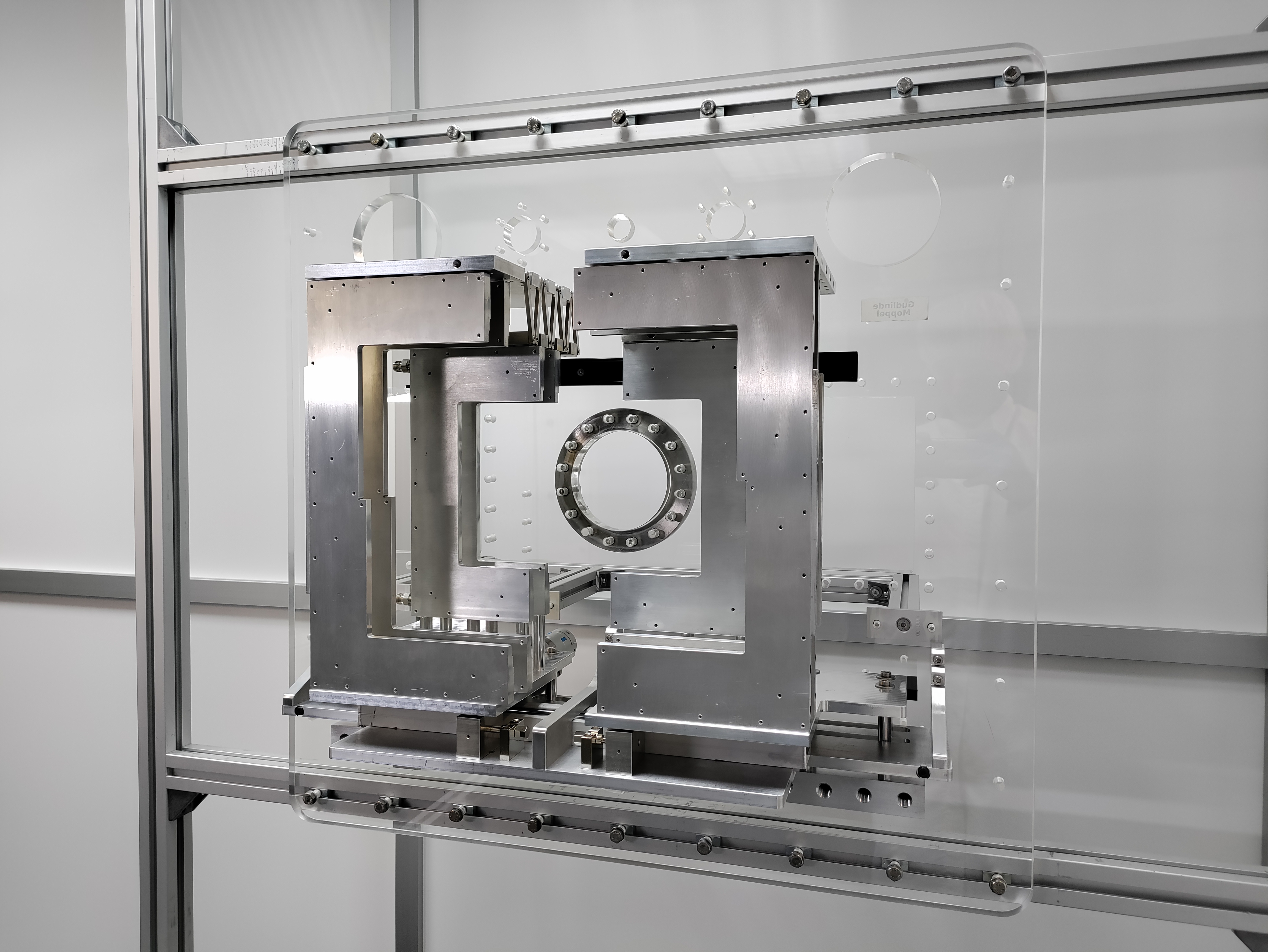}
            \caption{}
            \label{Fig:MVD_Mockup}
        \end{subfigure}
    \end{minipage}
    \hfill
    \begin{minipage}{.42\linewidth}
        \begin{subfigure}[t]{\linewidth}
            \centering
            \includegraphics[height=5cm]{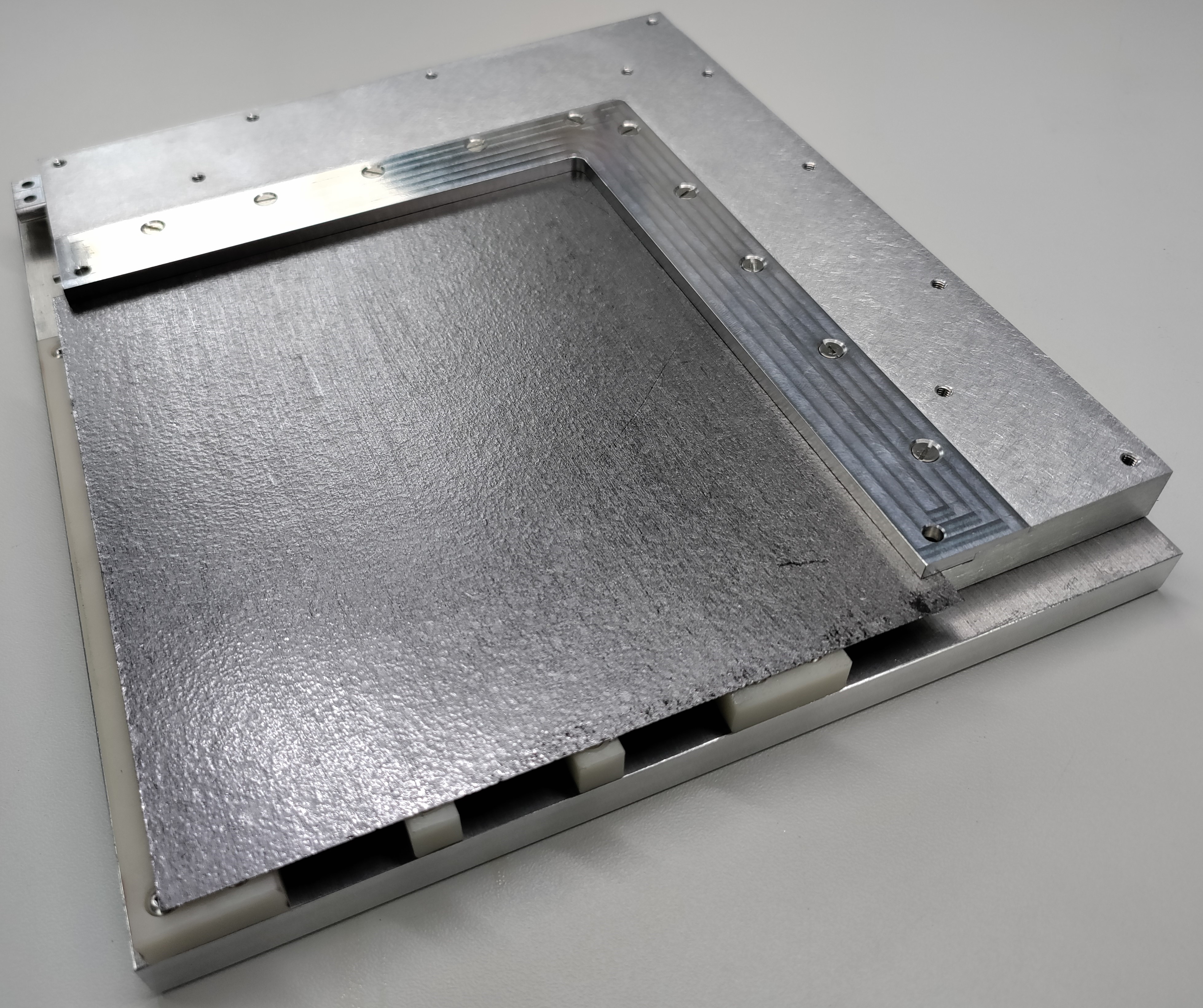}
            \caption{}
            \label{Fig:MVD_Jig}
        \end{subfigure}
    \end{minipage}
    \caption{The full-scale mechanical mock-up of the MVD~(\subref{Fig:MVD_Mockup}). TPG carrier mounted to a heat sink on the integration jig for bonding pad support.~(\subref{Fig:MVD_Jig}).}
\end{figure}

\paragraph{Module Assembly} The challenge of double-sided integration of large modules is met by a jig-less integration on a laser micro-structured TPG Carrier directly in the heat sink which has been explained in section~\ref{Ch:Carrier}. Additionally, a reworking strategy, analog to the TPG laser hatching technique, has been developed that allows to laser-ablate single sensors from the carrier if necessary. During integration, the Heat Sinks are mounted to a plate which also serves as local mechanical support during the full-automatic aluminum wire bonding of sensors to FPCs.
\vspace{-3pt}
\paragraph{Detector Assembly} Half-Stations are interconnected to Half-Detectors on Base- and actively cooled Top-Plates housing the LV distribution. They are placed on a Master Table comprising the Remote Positioning System with a rail system and pneumatic motors outside the vacuum to move the independent, rigid Half-Detectors between measuring and safe position. The assembly is being optimized in full-scale mechanical mock-up with final components on a PMMA flange~(figure~\ref{Fig:MVD_Mockup}).\par

\section{Outlook}
The integration concept of the MVD has been developed and tuned, preparing the project for pre-series production of modules in early 2025. The detector mechanics are nearly finalized and minor changes will be implemented based on the experience gained with a full-scale mock-up. A beam telescope based on MIMOSIS sensors will be operational at the mCBM test setup at GSI in 2025 to test the CBM-compatible readout and synchronization with other subsystems in a free-streaming readout. In parallel, a full-scale pre-production module is being assembled and validated.\par
The final sensor will be submitted mid 2025. Mass-testing will run in parallel to the design finalization and production of the detector electronics. The MVD will be assembled in the clean rooms at IKF Frankfurt and GSI to be installed in the cave for the CBM commissioning in 2028.
\vspace{-4pt}
\acknowledgments
The author acknowledges support from BMBF (Project-IDs 05P21RFFC2 and 05H24RF5), HFHF, GSI, and Eurizon.

\end{document}